\documentclass{article}

\usepackage{arxiv}

\usepackage[utf8]{inputenc} 
\usepackage[T1]{fontenc}    
\usepackage{hyperref}       
\usepackage{url}            
\usepackage{booktabs}       
\usepackage{amsfonts}       
\usepackage{nicefrac}       
\usepackage{microtype}      
\usepackage{lipsum}
\usepackage{graphicx}
\usepackage[bottom]{footmisc}
\usepackage[utf8]{inputenc}
\usepackage[T1]{fontenc}
\usepackage{multirow}
\usepackage{xr-hyper}
\usepackage[english]{babel}

\usepackage{hyperref, url, booktabs, amsfonts, nicefrac, microtype, lipsum, comment}

\usepackage{verbatim}
\usepackage{cite}
\usepackage{amsmath,amssymb,amsfonts}
\usepackage{algorithmic}
\usepackage{multicol, blindtext}
\usepackage{textcomp}
\usepackage[table,xcdraw]{xcolor}
\def\BibTeX{{\rm B\kern-.05em{\sc i\kern-.025em b}\kern-.08em
    T\kern-.1667em\lower.7ex\hbox{E}\kern-.125emX}}

\title{Towards future directions in data-integrative supervised prediction of human aging-related genes}

\author{
  Qi Li\textsuperscript{1}, Khalique Newaz\textsuperscript{1,2}, and Tijana Milenkovi\'{c}\textsuperscript{1,*}\\
  1. Department of Computer Science and Engineering, Lucy Family Institute for Data \& Society, \\
  and Eck Institute for Global Health (EIGH), University of Notre Dame, Notre Dame, IN 46556, USA.\\
  2. Center for Data and Computing in Natural Sciences (CDCS), Institute for Computational Systems Biology,\\
  Universit\"{a}t Hamburg, 20146, Hamburg, Germany.\\
  \texttt{qli8@nd.edu, khalique.newaz@uni-hamburg.de, and tmilenko@nd.edu} \\
}

\begin{document}
\maketitle

\begin{abstract}
\textbf{Motivation}: Identification of human genes involved in the aging process is critical due to the incidence of many diseases with age. A state-of-the-art approach for this purpose infers a weighted dynamic aging-specific subnetwork by mapping gene expression (GE) levels at different ages onto the protein-protein interaction network (PPIN). Then, it analyzes this subnetwork in a supervised manner by training a predictive model to learn how network topologies of known aging- vs. non-aging-related genes change across ages. Finally, it uses the trained model to predict novel aging-related genes. However, the best current subnetwork resulting from this approach still yields suboptimal prediction accuracy. This could be because it was inferred using outdated GE and PPIN data. Here, we evaluate whether analyzing a weighted dynamic aging-specific subnetwork inferred from newer GE and PPIN data improves prediction accuracy upon analyzing the best current subnetwork inferred from outdated data.\\
\textbf{Results}: Unexpectedly, we find that not to be the case. To understand this, we perform aging-related pathway and Gene Ontology (GO) term enrichment analyses. We find that the suboptimal prediction accuracy, regardless of which GE or PPIN data is used, may be caused by the current knowledge about which genes are aging-related being incomplete, or by the current methods for inferring or analyzing an aging-specific subnetwork being unable to capture all of the aging-related knowledge. These findings can potentially guide future directions towards improving supervised prediction of aging-related genes via -omics data integration. \\
\textbf{Availability and implementation}: All data and code are available at \url{https://nd.edu/~cone/DirectionsAging/}.\\
\textbf{Contact}: tmilenko@nd.edu
\end{abstract}

\section{Introduction} \label{sect:intro}
\subsection{Motivation and background} \label{sect:intro-motivation}

Human aging is a biological process associated with increased susceptibility to chronic disorders, such as cancer, cardiovascular, Parkinson's, and Alzheimer's disease \cite{uyar2020single, ferrucci2020measuring}. The aging process has a strong genetic basis, e.g., genomic instability or DNA somatic mutations \cite{rodriguez2011aging}. So, to understand it, treat associated diseases, and improve life quality for the elderly, it is critical to identify ``hallmark'' genes that drive the aging-related molecular mechanisms \cite{liguori2018oxidative}.

Traditionally, such ``hallmark'' genes have been identified by \emph{wet lab} experiments \cite{bolignano2014aging, paschos2012obesity}. These efforts have yielded valuable public knowledge about which genes are related to the aging process \cite{lu2004gene, berchtold2008gene, simpson2011microarray,tacutu2017human, jia2018analysis}. Yet, such knowledge is limited, because wet lab experimentation is difficult due to the ethical constraints and long life span of the human species \cite{emanuel2000makes}. So, by benefiting from the existing wet lab experimental aging-related knowledge plus wealth of recent -omics data, one can \emph{computationally} predict novel aging-related genes as those that share -omics ``signatures'' with the known wet lab experimental aging-related genes \cite{fabris2017review}. Unlike with wet lab experiments, computational analyses of the aging process can be done on a large, systems-level scale.

Such computational prediction of aging-related genes is typically carried out by supervised classification using gene expression (GE) data or protein-protein interaction network (PPIN) data \cite{fabris2017review}. In particular, GE-based approaches predict a gene as aging-related based on whether its expression level varies with age \cite{jia2018analysis, lu2004gene, berchtold2008gene, holtman2015induction, simpson2011microarray}. While GE-based approaches capture aging-specific information (i.e., the changes of gene expression with age), they ignore complex interactions between genes' protein products, which ultimately carry out all biological processes \cite{chen2014identifying}, including aging. This is why PPIN-based approaches have been introduced, which predict a gene as aging-related if its network topology is ``similar enough'' to the network topology of known aging-related genes \cite{freitas2011data, fang2013classifying, fabris2016extensive}. While PPIN-based approaches capture interactions between proteins, their interactions span different conditions, such as diseases or tissues. In our case, they span different ages, meaning that they do not capture aging-specific information. In other words, their considered PPIN data are context-unspecific (i.e., aging-unspecific). To address the above drawbacks, more recent studies have focused on predicting aging-related genes using both aging-specific GE data and entire context-unspecific PPIN data \cite{kerepesi2018prediction, faisal2014dynamic, Elhesha2019, li2020supervisedTCBB, li2021improved}.

Our group has pioneered this research direction by inferring an aging-specific subnetwork that captures both aging-specific information from GE data and interactions from PPIN data through a series of studies/methods \cite{faisal2014dynamic, li2019supervisedBIBM, newaz2020improving, li2020supervisedTCBB,li2021improved}. Our most recent finding is that inferring an aging-specific subnetwork that is both weighted and dynamic is superior (in terms of quality of aging-related gene predictions made from it) than inferring an aging-specific subnetwork that is unweighted or static as well as using entire context-unspecific PPIN data \cite{li2021improved}.

Intuitively, to infer such a weighted dynamic aging-specific subnetwork, we used network propagation \cite{Komurov2010, leiserson2015pan} to map gene expression levels from GE data onto a context-unspecific PPIN via random walks or diffusion. This resulted in assigning each interaction in the PPIN with an age-specific weight for each age present in the GE data. Such weighted interactions at a given age form a weighted age-specific subnetwork snapshot. The collection of age-specific snapshots in the increasing order of ages present in the GE data forms a weighted dynamic aging-specific subnetwork.
We believe that such a subnetwork results in higher-quality aging-related predictions than an unweighted or static subnetwork because the amounts of aging-specific information that different interactions carry are captured by their weight differences \emph{(1)} within each age, unlike in an unweighted subnetwork, and \emph{(2)} across different ages, unlike in a static subnetwork.

Briefly, we make aging-related gene predictions from a subnetwork as follows. Via several iterations \cite{li2019supervisedBIBM, li2020supervisedTCBB, li2021improved}, we developed a comprehensive framework that uses a variety of network features and classifiers in cross-validation. The framework relies on established knowledge, primarily from GenAge \cite{tacutu2017human}, about which genes are aging- vs. non-aging-related; this knowledge are genes' ground truth labels. The framework trains a predictive model (feature-classifier combination) on a part of the genes to learn network feature differences between the known aging- vs. non-aging-related genes from the training data. 
Then, it tests the model on the remaining genes, by examining how well it distinguishes between the known aging- vs. non-aging-related genes from the testing data. That is, the model predicts each gene from the testing data as either aging-related or not. Then, the model's accuracy, i.e., the quality of its predictions (whether the genes' predicted and ground truth labels match) is evaluated via the area under precision-recall curve (AUPR), precision, recall, and F-score. Finally, the most accurate of all predictive models is selected for the subnetwork of interest.

Nonetheless, even this newest (weighted and dynamic) aging-specific subnetwork \cite{li2021improved}, which is the state-of-the-art, yields suboptimal accuracy. Our postulations for this finding, which we explore in this study, are: 

\begin{enumerate}
    \item The GE data (from an over a decade old study \cite{berchtold2008gene}) and/or the PPIN data (from HPRD, which has not been updated for about a decade \cite{keshava2008human}) that we used to infer our weighted and dynamic aging-specific subnetwork are outdated.
    \item The current aging-related knowledge, i.e., ground truth labels from GenAge, might be incomplete or otherwise noisy and thus cannot reliably guide computational prediction of novel aging-related knowledge.
    \item The methods used to infer an aging-specific subnetwork or to predict aging-related genes from an inferred subnetwork \cite{li2019supervisedBIBM, li2020supervisedTCBB,li2021improved} cannot capture all of the current aging-related knowledge.
\end{enumerate}

Note that the reason why we used the older Berchtold GE and HPRD PPIN data in all of our previous studies is because our group started the research direction of inferring and analyzing aging-specific subnetworks in 2012, with the first paper published in 2014 \cite{faisal2014dynamic}. The only aging-specific GE data for human that encompassed multiple ages and enough samples for each age at the time was that curated by Berchtold \emph{et al.} via a microarray technology. Similarly, HPRD was a state-of-the-art human PPIN data at that time. Since then, we proposed a series of more advanced methods for inferring or analyzing aging-specific subnetworks \cite{newaz2020improving, li2019supervisedBIBM, li2020supervisedTCBB,li2021improved}. For fairness of evaluation of every new method against previous ones, we kept the input (i.e., GE and PPIN) data the same.
More recently, newer aging-specific GE data with enough ages and enough samples per age, produced using a next-generation sequencing technology (RNA-Seq), have become available, i.e., GE data from the Genotype-Tissue Expression (GTEx) project \cite{gtex2015genotype}. Also, newer, regularly updated PPIN data exist in the BioGRID database \cite{oughtred2021biogrid}.

In this study, we primarily (although not only) examine whether using the newer GTEx GE or BioGRID PPIN data to infer a weighted dynamic aging-specific subnetwork will improve the quality of predicted aging-related genes compared to using the older Berchtold GE and HPRD PPIN data. 

\begin{figure}[!ht]
\centering
  \includegraphics[width= \linewidth]{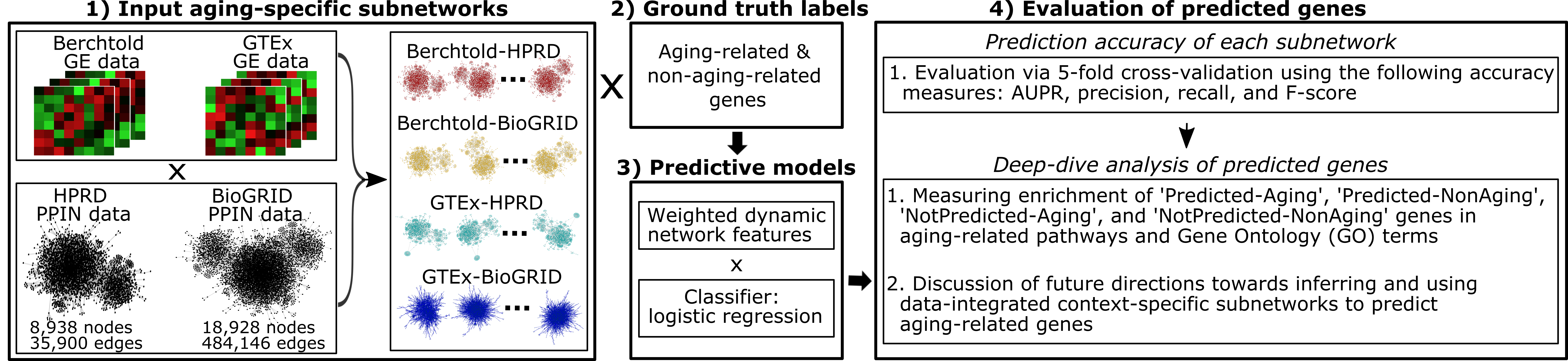}
  \vspace{-0.5cm}
  \caption{Summary of our study. See the text for details.}
  \label{fig:workflow}
  \vspace{-0.2cm}
\end{figure}

\subsection{Our study and contribution}\label{sect:intro-ourstudy}

We summarize our study in Fig.~\ref{fig:workflow}. To properly examine which of the two data components (GE data or PPIN data) might lead to better prediction accuracy, we construct four weighted dynamic aging-specific subnetworks by varying one data component at a time, as shown in Table \ref{table:netproperty}. To construct and analyze (i.e., make predictions from) the four subnetworks, we rely on our well-established framework discussed above \cite{li2020supervisedTCBB, li2021improved}.

\begin{table}[!ht]
    \centering
    \caption{Our four weighted dynamic aging-specific subnetworks (in bold) constructed from the four combinations of GE and PPIN data.}
    \label{table:netproperty}
    \vspace{-0.2cm}
    \begin{tabular}{|l|c|c|c|c|}
      \hline
      PPIN \hspace{0.2cm} \textbackslash  \hspace{0.2cm} GE data      & Berchtold (old)   & GTEx (new)   \\ \hline
      HPRD (old)          &  \textbf{Berchtold-HPRD}      & \textbf{GTEx-HPRD}            \\ \hline
      BioGRID (new)      &  \textbf{Berchtold-BioGRID}   & \textbf{GTEx-BioGRID}         \\ \hline
    \end{tabular}
\end{table}

Given our four subnetworks, postulation 1 would be validated if using GTEx-BioGRID is better than using all of Berchtold-HPRD, Berchtold-BioGRID, and GTEx-HPRD. Shockingly, we find this not to be the case. That is, generally, neither using newer PPIN data nor using newer GE data helps. To attempt to understand this finding, we perform several additional analyses, including the following.

All four subnetworks perform statistically significantly better than expected by chance, i.e., each subnetwork captures some of the aging-related knowledge. So, we aim to test whether the different subnetworks yield complementary or duplicated predictions. When we look into the overlaps of their predicted aging-related genes, we find that the different subnetworks are overall complementary to each other, with relatively low (although statistically significant) average pairwise overlaps of 34\%. Hence, even though newer GE or PPIN data do not necessarily improve upon their older counterparts, they do capture aging-related knowledge that older data cannot.

However, our subnetworks are imperfect regardless of what data they are inferred from: the highest AUPR, precision, recall, and F-score over all subnetworks are 50\%, 67\%, 45\%, and 52\%, respectively. To understand such suboptimal performance, we examine potential underlying reasons. In particular, we rely on existing data on human proteins' memberships in aging-related KEGG pathways \cite{kanehisa2021kegg} and their Gene Ontology (GO) term annotations \cite{gene2021gene}. We test the enrichment of four groups of aging- and non-aging-related genes in the aging-related pathways and GO terms. These four gene groups are presented in Table \ref{table:genegroup}.

\begin{table}[!ht]
    \centering
    \caption{The four considered groups of aging- and non-aging-related genes. The gene groups are named based on a combination of their predicted label and their ground truth label.}
    \label{table:genegroup}
    \vspace{-0.2cm}
    \begin{tabular}{|cc|cc|}
    \hline
    \multicolumn{2}{|c|}{\multirow{2}{*}{}}   & \multicolumn{2}{c|}{Ground truth label}  \\ \cline{3-4} 
    \multicolumn{2}{|c|}{}   & \multicolumn{1}{c|}{Aging-related}  & Non-aging-related   \\ \hline
    \multicolumn{1}{|c|}{\multirow{2}{*}{\begin{tabular}[c]{@{}c@{}}Predicted as\\ aging-related?\end{tabular}}} & Yes     &
    
    \multicolumn{1}{c|}{\textbf{Predicted-Aging}}    & \textbf{Predicted-NonAging}    \\ \cline{2-4} 
    
    \multicolumn{1}{|c|}{}   & No & \multicolumn{1}{c|}{\textbf{NotPredicted- Aging}} & \textbf{NotPredicted-NonAging} \\ \hline
    \end{tabular}
\end{table}

Predicted-Aging genes, being supported by both our subnetworks and GenAge, are our positive control and should be enriched in aging-related pathways or GO terms. Similarly, NotPredicted-NonAging genes, being supported by neither our subnetworks nor GenAge, are our negative control and should not be enriched in any aging-related pathways and GO terms. Indeed, as expected, we find that Predicted-Aging genes are enriched in almost all aging-related pathways and GO terms, and NotPredicted-NonAging genes enriched in none of them. Then, we examine the enrichment of Predicted-NonAging and NotPredicted-Aging genes in the aging-related pathways and GO terms, for the following reasons.

Our suboptimal precision means that a portion of each subnetwork's predictions are currently not known to be aging-related; these correspond to Predicted-NonAging genes. If these genes that are currently missing from GenAge are enriched in aging-related pathways or GO terms and are thus more similar to the positive control genes than the negative control genes, then this could imply that postulation 2 holds, i.e., that GenAge is incomplete. In this case, Predicted-NonAging genes could be considered as highly ranked candidates for future wet lab validation. 
Indeed, we find this to be the case -- Predicted-NonAging genes are statistically significantly enriched in multiple aging-related pathways or GO terms.

Our suboptimal recall means that each subnetwork fails to predict a portion of known aging-related genes from GenAge; these correspond to NotPredicted-Aging genes. 
If these genes are enriched in aging-related pathways or GO terms, it could mean that the aging process has at least two distinct network ``signatures'': one signature shared between those genes from GenAge that can be captured by our subnetworks, and a different signature (or signatures) that the remaining genes from GenAge have but that our subnetworks cannot recognize. This would correspond to postulation 3, i.e., network methods used to infer a weighted dynamic aging-specific subnetwork or to predict aging-related genes from an inferred subnetwork being unable to capture all of the current aging-related knowledge.
Otherwise, if NotPredicted-Aging genes are not enriched in aging-related pathways or GO terms, this would mean that there might be discrepancies between GenAge aging-related ground truth data and aging-related pathway and GO term data, possibly because GenAge is noisy or because the different databases capture complementary aging-related knowledge. We find the former to be the case --  NotPredicted-Aging genes are statistically significantly enriched in multiple aging-related pathways or GO terms.
If indeed postulation 3 holds, i.e., if our current network construction or analysis approaches fail to capture all of the current aging-related knowledge, then a solution would be to revise and redesign these approaches in hope of capturing the aging-related knowledge better. This is the subject of future work. 

\vspace{-0.1cm}
\section{Results} \label{sect:results}
\vspace{-0.1cm}
As already discussed, we consider four weighted dynamic aging-specific subnetworks: Berchtold-HPRD, Berchtold-BioGRID, GTEx-HPRD, and GTEx-BioGRID. We  test whether GTEx-BioGRID, which is inferred using both newer GE and PPIN data, outperforms the remaining three subnetworks that are inferred using older GE data or older PPIN data (or both). 
To test this, first, for each subnetwork, we consider nine predictive models (i.e., feature-classifier combinations). We train and test each predictive model via 5-fold cross-validation in terms of prediction accuracy (AUPR, precision, recall, and F-score averaged over the five folds). Then, we select the best predictive model that yields the highest AUPR, to give each subnetwork the best-case advantage (Section \ref{sect:result-predmodel}). For fairness, we force the gene sets that are randomly split into the training and testing data for cross-validation to be same for all predictive models across all four subnetworks (see Section \ref{sect:method-data-gt} for details). 
Second, we compare the selected best predictive models to evaluate which subnetwork results in the highest prediction accuracy (Section \ref{sect:result-cv}). Third, we analyze whether the four subnetworks are predicting redundant or complementary aging-related genes (Section \ref{sect:result-overlap}). Finally, we present a deep-dive analysis in terms of enrichment of the predicted genes in aging-related pathways and GO terms (Section \ref{sect:result-deep}).

\subsection{Selecting the best predictive model for each subnetwork \label{sect:result-predmodel}}
The considered nine predictive models per subnetwork are combinations of nine features and one classifier. We consider the nine best features among 30 features evaluated in our previous study \cite{li2021improved} (Supplementary Section S1.2). We use logistic regression as the classifier for all predictive models because it consistently performed the best among nine prominent classifiers evaluated in our previous studies \cite{li2019supervisedBIBM, li2020supervisedTCBB, li2021improved}.  

For each of the four subnetworks, all nine predictive models perform statistically significantly better than expected by chance (adjusted $p$-values $< 0.05$), with respect to all four accuracy measures (Supplementary Tables S1-S4 and Supplementary Figs. S1-S4). For each subnetwork, the best-performing predictive model brings sufficient (although not always statistically significant) improvement compared to the remaining eight predictive models. This is why we choose the best-performing predictive model for a given subnetwork.

Overall, it is typically different predictive models that are selected for the different subnetworks. That is, the four subnetworks yield three distinct best predictive models. The same best predictive model is selected only for the two HPRD-based subnetworks (Berchtold-HPRD and GTEx-HPRD).

\subsection{Comparing prediction accuracy of the four subnetworks}\label{sect:result-cv}
Given the selected best predictive model for each of the four subnetworks, we compare the prediction accuracy of the four subnetworks. With this, we aim to test postulation 1: whether using GTEx-BioGRID that is inferred from both newer GE and PPIN data would outperform all other three subnetworks that are inferred from older GE data or older PPIN data (or both). 

\begin{figure}[!ht]
\centering
  \includegraphics[width= 0.6\linewidth]{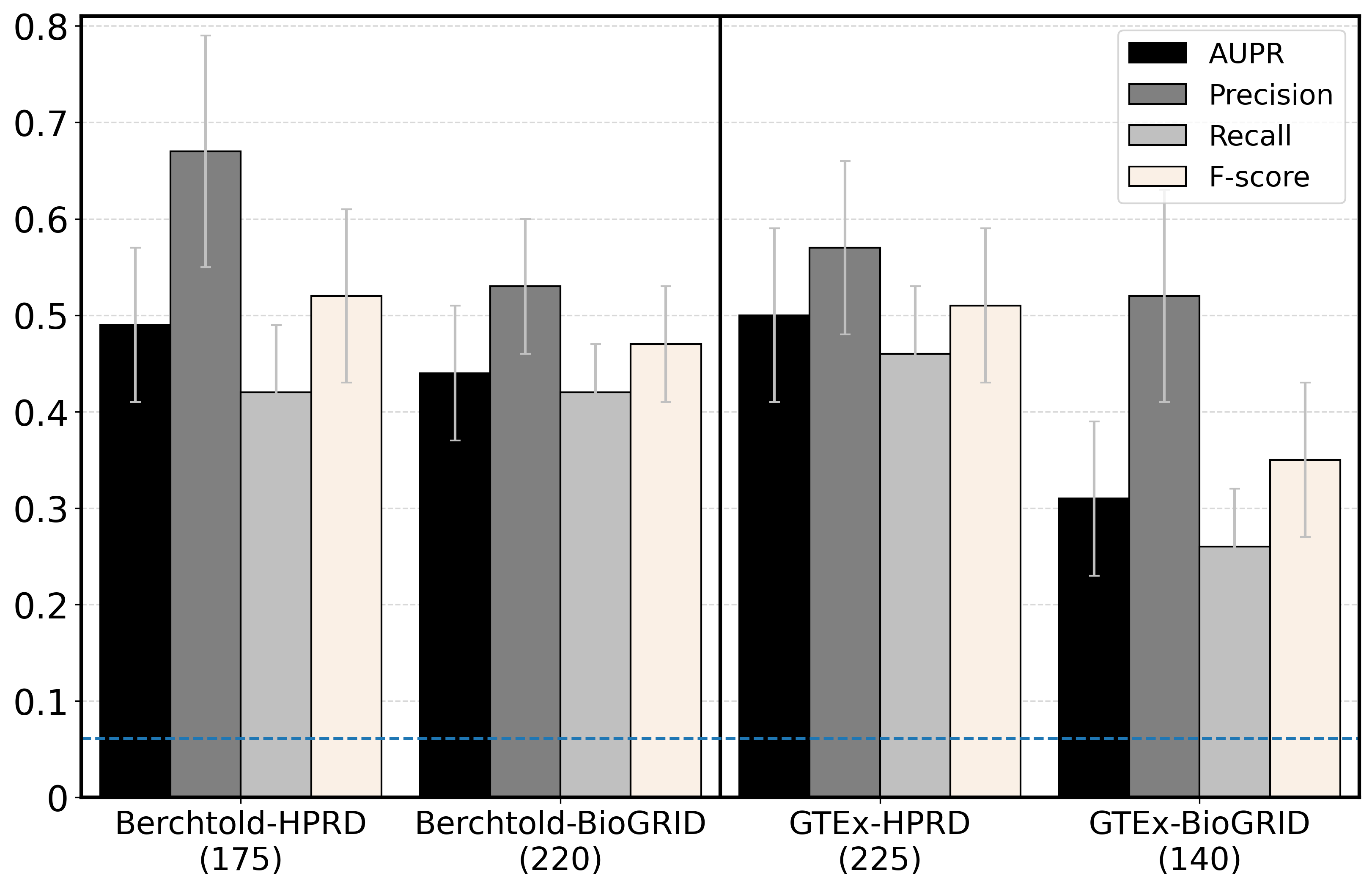}
  \vspace{-0.3cm}
  \caption{The prediction accuracy in terms of AUPR, precision, recall, and F-score of the four weighted dynamic aging-specific subnetworks, each under its best predictive model. The number below each subnetwork name represents the number of genes that are predicted as aging-related by the corresponding subnetwork. The blue dashed line indicates the prediction accuracy scores expected by chance, i.e., the fraction of all genes in the ground truth data that are labeled as aging-related.}
  \label{fig:APRF}
  \vspace{-0.3cm}
\end{figure}

First, we ask whether using newer BioGRID PPIN data is better than using its older HPRD counterpart when the GE data is fixed, i.e., whether Berchtold-BioGRID improves upon Berchtold-HPRD and whether GTEx-BioGRID improves upon GTEx-HPRD. We find neither of the two to hold (Fig. \ref{fig:APRF}): Berchtold-BioGRID performs marginally worse than Berchtold-HPRD in terms of all four prediction accuracy measures, and GTEx-BioGRID performs marginally worse than GTEx-HPRD in terms of all four prediction accuracy measures.

Second, we ask whether using newer GTEx GE data is better than using its older Berchtold counterpart when the PPIN data is fixed, i.e., whether GTEx-HPRD improves upon Berchtold-HPRD and whether GTEx-BioGRID improves upon Berchtold-BioGRID. We find that these do not necessarily hold (Fig. \ref{fig:APRF}). Namely, GTEx-HPRD performs marginally worse than Berchtold-HPRD in terms of precision and F-score, although marginally better in terms of AUPR and recall. GTEx-BioGRID performs marginally worse than Berchtold-BioGRID for all four prediction accuracy measures. 

Overall, GTEx-BioGRID that is inferred from both newer GE and PPIN data does not outperform any subnetwork inferred from at least one older data component. That is, postulation 1 does not hold. In fact, Berchtold-HPRD that is inferred from both older GE and PPIN data marginally outperforms all other subnetworks in terms of precision and F-score. Meanwhile, Berchtold-HPRD is the second-best subnetwork in terms of AUPR and recall. 

\subsection{Examining prediction overlaps between the subnetworks \label{sect:result-overlap}} 
Thus far, we have found that using both newer GE and PPIN data does not necessarily improve the prediction accuracy upon using older GE data or older PPIN data (or both); also, all four subnetworks yield prediction accuracies that are statistically significantly better than at random. Given this, we ask whether our four subnetworks are predicting complementary or redundant aging-related knowledge by examining overlaps of their predictions. Namely, we split predicted aging-related genes of each subnetwork into two groups: true positives (i.e., genes predicted as aging-related and labeled as aging-related in the ground truth data) and novel predictions (i.e., genes predicted as aging-related but are currently labeled as non-aging-related in the ground truth data); the former is more confident than the latter to actually be aging-related.

\begin{figure}[!ht]
\centering
  \vspace{-0.2cm}
  \includegraphics[width= 0.6\linewidth]{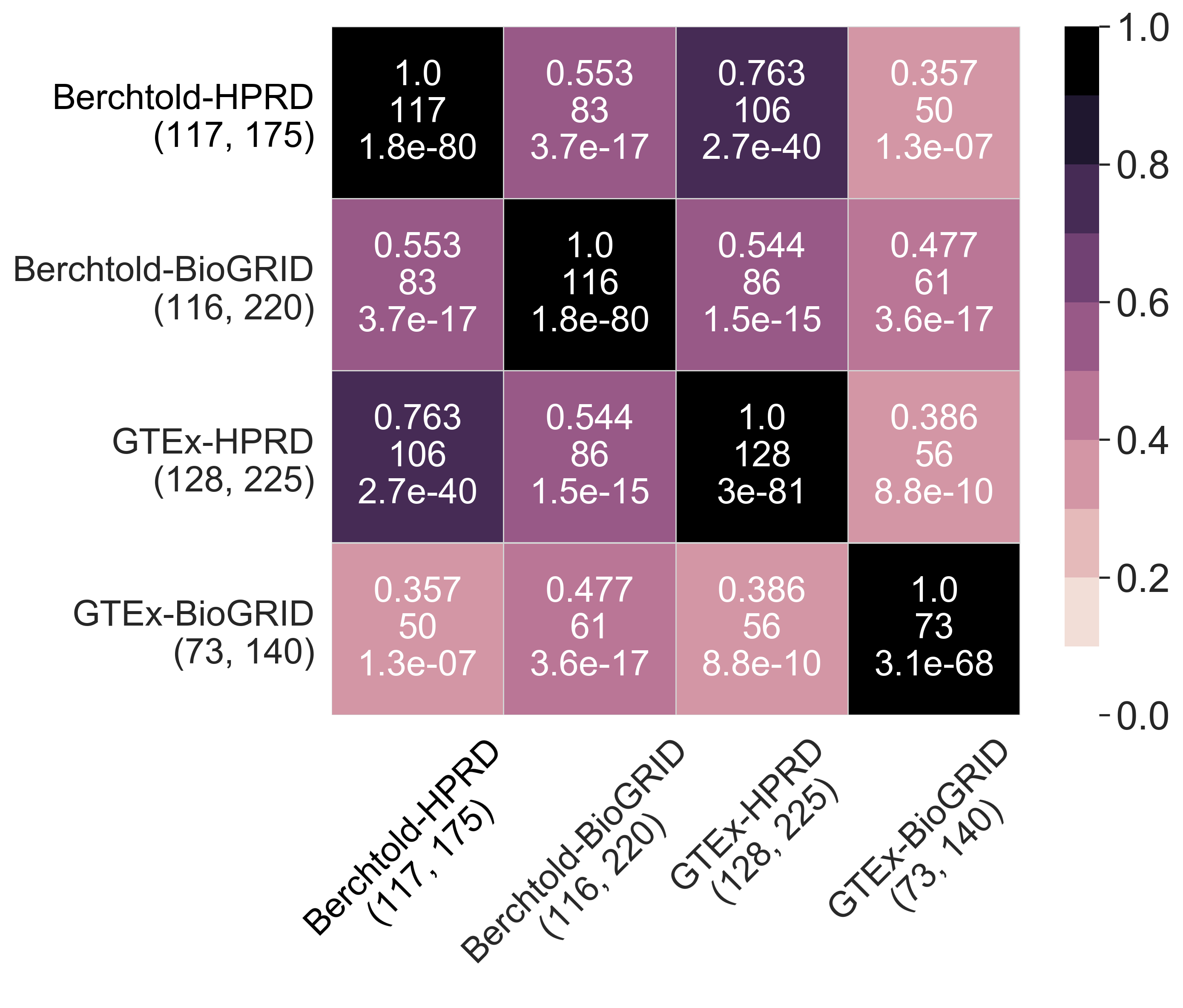}
  \vspace{-0.5cm}
  \caption{Pairwise overlaps in terms of Jaccard indices of true positives for each pair of the four considered subnetworks. By true positives, we mean genes that are predicted as aging-related and are also present in GenAge. The two numbers in the parenthesis below each subnetwork name represent the number of true positives and the number of all predicted aging-related genes for the given subnetwork, respectively. In a cell, corresponding to a pair of subnetworks, the three numbers represent the Jaccard index (top), the raw number of genes in the overlap (middle), and the adjusted $p$-value indicating whether the overlap is statistically significantly high. The color shades are driven by Jaccard indices, where a darker color means a higher Jaccard index. For overlaps of novel predictions (rather than true positives), see Supplementary Fig. S6.}
  \label{fig:overlap}
  \vspace{-0.3cm}
\end{figure}

For true positive predictions, we find that all pairwise overlaps are statistically significantly high (i.e., adjusted $p$-values $< 0.05$) but in reality still quite complementary. Namely, the largest, smallest, and average Jaccard indices of all pairwise overlaps are $76.3\%, 35.7\%,$ and $51.3\%$, respectively (Fig. \ref{fig:overlap}). Moreover, Berchtold-HPRD, Berchtold-BioGRID, GTEx-HPRD, and GTEx-BioGRID capture $4.3\%$, $11.2\%$, $7.8\%$, and $5.5\%$ of GenAge-based aging-related knowledge that none of the other three subnetworks capture, respectively.

Results for novel predictions are qualitatively similar to the results for true positive predictions, i.e., all overlaps are statistically significantly high but in reality still quite complementary (Supplementary Fig. S6). Namely, the largest, smallest, and average Jaccard indices of all pairwise overlaps are $42.2\%, 5.9\%,$ and $16.9\%$, respectively. Moreover  Berchtold-HPRD, Berchtold-BioGRID, GTEx-HPRD, and GTEx-BioGRID capture $17.2\%$, $57.7\%$, $40.2\%$, and $52.2\%$ of novel aging-related knowledge that none of the other three subnetworks capture, respectively.

So, even though using newer GE or PPIN data does not improve upon using their older counterparts, the former does capture aging-related knowledge that using older data does not. 

\subsection{Validating predictions using aging-related pathways and GO terms \label{sect:result-deep}}
Although all four subnetworks yield statistically significantly high prediction accuracies, their accuracies are suboptimal, i.e., the highest accuracy score is ``only'' $67\%$. To understand such suboptimal performance, we rely on existing data on human proteins' memberships in aging-related KEGG pathways and their aging-related GO term annotations. As typically done, we analyze whether our predictions are statistically significantly enriched in aging-related pathways or GO terms. To do so, we split the genes from the ground truth data into four groups (Predicted-Aging, Predicted-NonAging, NotPredicted-Aging, and NotPredicted-NonAging genes, as shown in Table \ref{table:genegroup}) based on whether their predicted labels agree with their ground truth labels (Section \ref{sect:intro-ourstudy}). Then, we test the enrichment of each gene group in aging-related pathways and GO terms. 
We use all five established aging-related pathways \cite{yu2021key} that annotate sufficiently many (at least three) of our ground truth labeled genes. We also use all 13 aging-related GO terms that annotate at least three genes in our ground truth data. 

\begin{figure}[!t]
\centering
  \vspace{-0.4cm}
  \includegraphics[width= 0.45\linewidth]{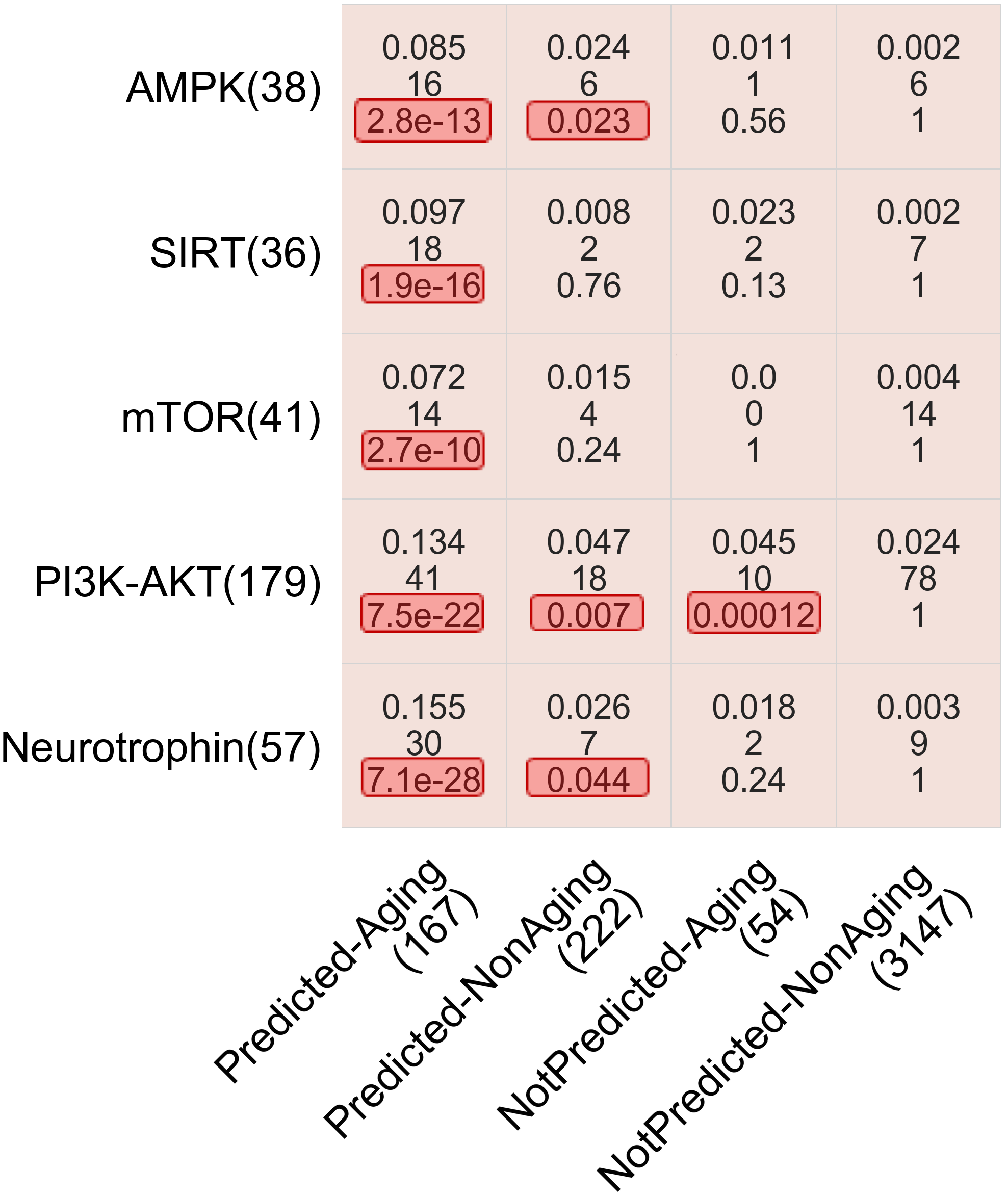}
  \vspace{-0.2cm}
  \caption{Enrichments of the four gene groups ($x$-axis) in the five aging-related pathways ($y$-axis). The number below a pathway name or a gene group name represents the gene count in the pathway or gene group. In each cell, the three numbers represent the overlap size as measured by the Jaccard index (top), the raw number of genes in the overlap (middle), and the adjusted $p$-value indicating whether the overlap size is statistically significantly high, i.e., whether the given gene group is statistically significantly enriched in the given pathway. The adjusted $p$-values below 0.05 are highlighted in red. Analogous results for aging-related GO terms are shown in Supplementary Fig. S7.}
  \label{fig:pathway}
  \vspace{-0.3cm}
\end{figure}

We find that the Predicted-Aging genes are significantly enriched (adjusted $p$-values $<0.05$) in all aging-related pathways (Fig. \ref{fig:pathway}) and over half of the aging-related GO terms (Supplementary Fig. S7). This is expected, as Predicted-Aging genes are supported as aging-related by both our subnetworks and the ground truth data and are thus our positive control (i.e., genes that are the most confident to be aging-related). 

On the hand, NotPredicted-NonAging genes are not significantly enriched (adjusted $p$-values $=1.0$) in any aging-related pathway or GO term (Fig. \ref{fig:pathway} and Supplementary Fig. S7). This is also expected, as  NotPredicted-NonAging genes are not supported as aging-related by either our subnetworks or the ground truth data and are thus our negative control (i.e., genes that are the most confident to be non-aging-related).

Next, we look into Predicted-NonAging, the gene group that causes suboptimal precision. We find this group of genes to be  significantly enriched (adjusted $p$-values $<0.05$) in three aging-related pathways (AMPK, PI3K-AKT, and Neurotrophin, Fig. \ref{fig:pathway}). That is, statistically significantly many Predicted-NonAging genes are linked to the aging process. In other words, postulation 2 seems to hold, i.e., GenAge appears to be incomplete. So, the novel predictions from our subnetworks could potentially guide future wet lab experiments to expand the current GenAge aging-related ground truth data. 

Finally, we look into NotPredicted-Aging, the gene group that causes suboptimal recall. We find this group of genes to be  significantly enriched in an aging-related pathway (PI3K-AKT, Fig. \ref{fig:pathway}) and two aging-related GO terms (GO:0007568--aging and GO:0008340--determination of adult lifespan, Supplementary Fig. S7). That is, these GenAge genes that are not captured by our subnetworks are indeed aging-related. This suggests that there may be multiple network ``signatures'' of aging-related genes: one that is shared between GenAge aging-related genes and our subnetworks, and other signature(s) that are not recognized by our subnetworks. In other words, postulation 3 seems to hold, i.e., our subnetworks or network features might have failed to recognize all network ``signatures'' of GenAge aging-related genes.

In summary, these analyses show that the enrichments of Predicted-Aging, Predicted-NonAging, and NotPredicted-Aging genes in aging-related pathways and GO terms are statistically significantly high, i.e., these are confident aging-related genes. On the other hand, the enrichments of NotPredicted-NonAging genes are not statistically significant, i.e., these are confident non-aging-related genes.

\vspace{-0.1cm}
\section{Discussion}\label{sect:discuss}
\vspace{-0.1cm}
In this study, we systematically evaluate whether GTEx-BioGRID that is inferred from both newer GE and PPIN data, improves the prediction accuracy upon its counterparts inferred from at least one older data component. We discuss our findings and provide future directions in this section.

\subsection{The choice of best predictive model matters}\label{sect:discuss-model}

The choice of predictive model for a given subnetwork matters (Section \ref{sect:result-predmodel}). The fact that no model performs the best for all subnetworks stresses the need to test multiple models, to give each subnetwork the best-case advantage. 

Also, given the fact that the four subnetworks are predicting complementary aging-related knowledge (Section \ref{sect:result-overlap}), it might be worth to pursue development of a novel network feature or an ensemble learning approach that would integrate the complementary aspect of the four subnetworks. 

\vspace{-0.1cm}
\subsection{Higher-quality PPIN or GE data might be needed} \label{sect:discuss-needdata}

When GE data is fixed, subnetworks inferred using newer BioGRID PPIN data never outperform subnetworks inferred using older HPRD PPIN data (Section \ref{sect:result-cv}). The superiority of using HPRD PPIN data may be due to HPRD being manually curated (i.e., ``read, analyzed, and interpreted by expert biologists'' according to the HPRD website), thus only containing interactions that are confident \cite{lazareva2021limits}, although among fewer genes than newer genes than newer PPIN data. 
On the other hand, PPIN databases that are not manually curated, including BioGRID, could potentially introduce technical or literature biases when newer interactions are added \cite{lazareva2021limits}. That is, those proteins that are used as ``baits'' for discovering new interactions \cite{lazareva2021limits},  have known biological functions,  are related to common diseases, etc., along with their interactions, may be studied more than the remaining proteins and their interactions. Henceforth, newer, regularly updated PPIN databases such as BioGRID might continuously increase their bias towards uncovering biology primarily about the aforementioned types of proteins and their interactions.

When PPIN data is fixed, subnetworks inferred using newer GTEx GE data  do not necessarily outperform subnetworks inferred using older Berchtold GE data  (Section \ref{sect:result-cv}).
This may be because of a variation between Berchtold and GTEx GE data: the samples in Berchtold GE data are grouped across 37 specific ages in the $[20-99]$ range, while the samples in GTEx GE data are grouped across six age groups (20-29, 30-39, 40-49, 50-59, 60-69, 70-79). So, the resulting Berchtold-based subnetworks have 37 snapshots while the GTEx-based subnetworks have six snapshots. To test whether the differences in the number or nature of the snapshots between Berchtold-based and GTEx-based subnetworks yield differences in their  accuracy, we further infer two subnetworks, as follows. First, we group samples in Berchtold GE data into the same six age groups as in GTEx GE data. Then, we integrate the modified Berchtold GE data with each of HPRD and BioGRID PPIN data, to infer two new subnetworks, each with six snapshots; we refer to these as Berchtold-HPRD-6 and Berchtold-BioGRID-6, respectively. This way, both the number and nature of snapshots are matched between Berchtold-HPRD-6 and GTEx-HPRD, as well as between Berchtold-BioGRID-6 and GTEx-BioGRID. Thus, comparison may be more fair between Berchtold-HPRD-6 and GTEx-HPRD  than   between Berchtold-HPRD and GTEx-HPRD. Similarly,  comparison may be more fair between Berchtold-BioGRID-6 and GTEx-BioGRID than between Berchtold-BioGRID and GTEx-BioGRID. 

When we compare Berchtold-HPRD-6 and GTEx-HPRD, as well as Berchtold-BioGRID-6 and GTEx-BioGRID, i.e., even when the age groups match between Berchtold and GTEx GE data, using newer GTEx GE and BioGRID PPIN data still does not improve accuracy compared to using at least one older data component (Supplementary Fig. S5). That is, GTEx-BioGRID is still inferior among all four subnetworks with six age groups (i.e., snapshots). In fact, Berchtold-HPRD-6 inferred using both older GE and  PPIN data performs marginally better than the other three six-snapshot subnetworks. Therefore, the difference in the nature of Berchtold and GTEx GE data does not account for why newer data does not help.

A side observation of this analysis is as follows. By comparing Berchtold-HPRD vs. Berchtold-HPRD-6 (both of which happen to be the best-performing networks in their respective evaluations), we can evaluate the effect of using specific ages vs. using age groups, respectively, to construct aging-specific subnetworks. We find that the two subnetworks yield almost indistinguishable prediction accuracies (Supplementary Fig. S5). Thus, the aforementioned effect seems to be minimal, i.e., it does not seem to matter whether specific ages or age groups are used. 
The implications of this result are as follows. When in GE data the samples are already provided as being grouped across age groups, without having information about their specific ages, one clearly has to use age groups.
On the other hand, when in GE data the samples' specific ages are known, one could use the specific ages directly or they could instead first group the ages into age groups and then use the groups. For the GE data we have evaluated, the choice between the two does not seem to matter in terms of accuracy. But the same might not necessarily hold for different GE data. So, ideally, one should empirically evaluate whether it is better to use specific ages or somehow form age groups.
However, it is unclear how exactly to define age groups, e.g., whether one should one should define an age group (i.e., construct an age group-specific snapshot) for every 5, 10, or more years of the human lifespan, as well as whether age groups (i.e., snapshots) should span age intervals of equal length or of differing lengths \cite{newaz2020network}. A possible way to address these challenges could be to start with the network snapshots corresponding to  specific ages and then use algorithms such as SCOUT \cite{hulovatyy2016scout} to computationally identify age groups. Namely, SCOUT is able to identify $n$ time points (in our context, ages) in the entire time interval where network structure (with respect to some network structural property, in SCOUT's case, community or clustering structure) significantly changes. This corresponds to identifying $n+1$ temporal segments (in our context, age groups) that are separated by these $n$ changing time points  and the corresponding $n+1$ network snapshots. The intuition here is that specific ages (that are consecutive in GE data) which have similar community (or other network) structures are likely to have similar biological ``signatures'' and hence should belong to the same age group.  See the SCOUT paper \cite{hulovatyy2016scout} for more details.

While it is still unclear why newer GTEx GE data does not outperform older Berchtold GE data, the former does offer opportunities that the latter does not. While Berchtold GE data only encompasses samples from the brain, GTEx GE data encompasses samples from 31 organs. This, combined with the fact that GTEx-based subnetworks perform significantly better than expected by chance and only marginally worse than Berchtold-based subnetworks, means that GTEx GE data allows for studying aging in tissues other than the brain. 

Finally, we comment on an additional opportunity of GE data. GTEx GE data was curated using RNA-seq in different \emph{tissues}. More recently, even newer biotechnology, single-cell RNA-seq, has allowed for curation of the Human Cell Atlas data that allows for even more detailed investigations of the human aging process -- in different \emph{cell types} \cite{uyar2020single}.

\subsection{More complete aging-related ground truth data might be needed} \label{sect:discuss-needgt} 

Of Predicted-Aging, Predicted-NonAging, and NotPredicted-Aging genes, which are supported by our subnetworks or ground truth data, only Predicted-Aging and Predicted-NonAging genes are captured (predicted as aging-related) by our subnetworks (Section \ref{sect:result-deep}). This implies that these two gene groups have similar network topologies (features) in our subnetworks. 
We aim to illustrate this by visualizing these genes' features in 2-dimensional (2D) vector space as described in Supplementary Section S1.4. Their features are indeed close in the 2D space, i.e., are topologically similar to each other in our subnetworks (Fig. \ref{fig:2d-berchtold-biogrid} and Supplementary Figs. S8-S11). 
The 2D visualization further supports postulation 2, i.e., GenAge might be incomplete. This may not be surprising, because human genes in GenAge are  orthologs of aging-related genes in model species. That is, GenAge  encompasses only the aging-specific ``biology'' that is common to the human species and model species. However, each species, especially human, has its unique aging-specific ``biology'' that is not shared with other species \cite{bronikowski2011aging}.

Thus, there is a need for more complete, human-specific knowledge on which genes are linked to aging. To the best of our understanding, the only such data come from GE studies such as the Berchtold study \cite{berchtold2008gene}, GTEx project \cite{jia2018analysis}, or the Human Cell Atlas \cite{uyar2020single}. But we already use such GE data to infer our subnetworks in the first place. Therefore, we cannot use that same data as the ground truth knowledge on which genes are linked to aging as well, as this would yield a circular argument. 
A potential solution for this challenge is to combine the genes in GenAge with those that are members of the aging-related pathways or are annotated by aging-related GO terms into a more comprehensive aging-related ground truth gene set. 
Another potential solution is to rely on predicted novel aging-related genes from existing computational studies, e.g., those predicted via network clustering \cite{hulovatyy2016scout}, network alignment \cite{faisal2014global}, unsupervised analyses of aging-specific subnetworks \cite{newaz2020improving}, or  supervised analyses of aging-specific subnetworks. In this study, it would mean adding Predicted-NonAging genes to the ground truth knowledge, as these can be viewed as confident candidates for future wet lab validation.

\vspace{-0.2cm}
\subsection{Advanced algorithms for network inference or analysis might be needed} \label{sect:discuss-needmethod}

\begin{figure}[!ht]
\centering
  \vspace{-0.2cm}
  \includegraphics[width=0.45\linewidth]{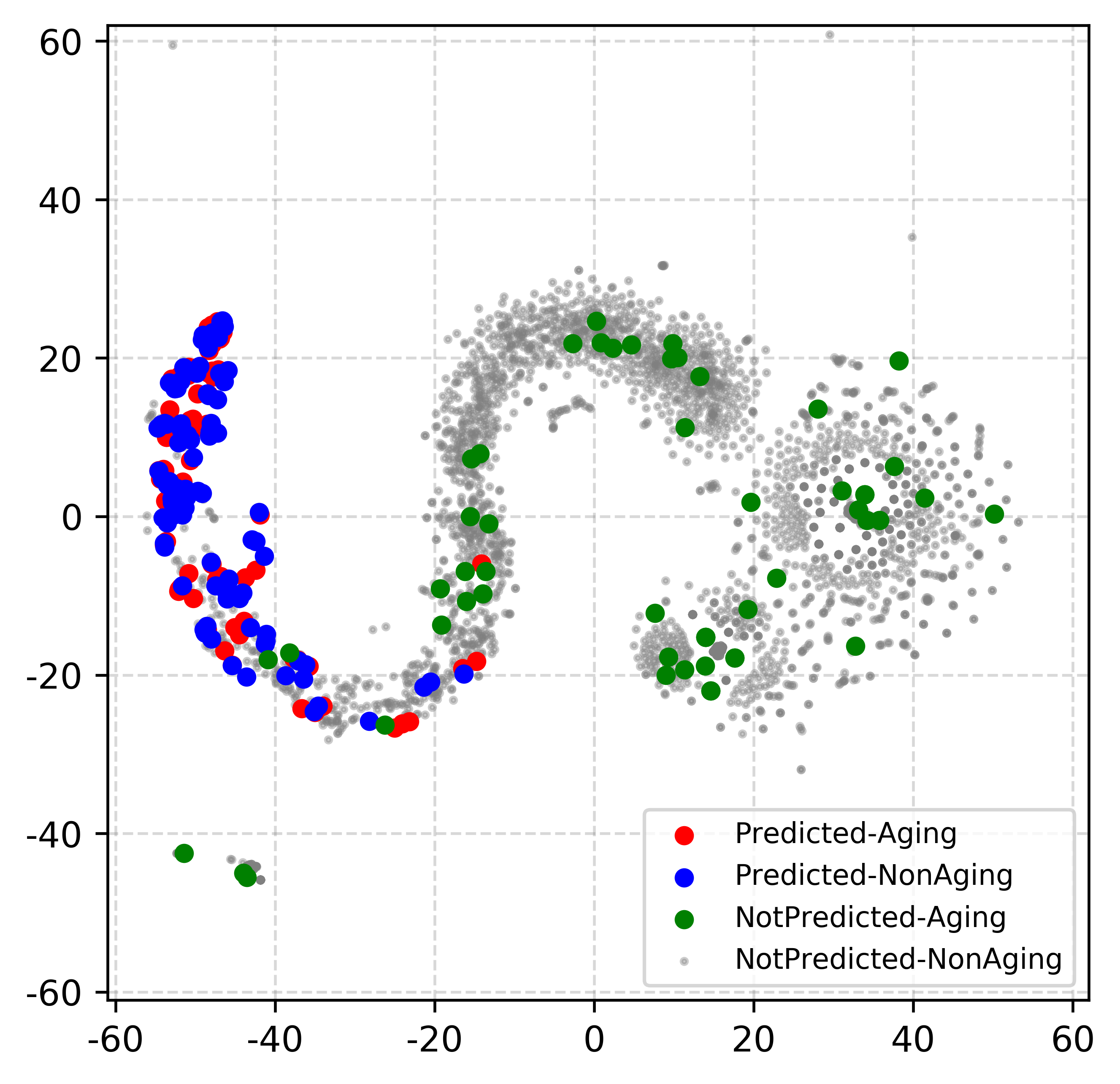}
  \vspace{-0.35cm}
  \caption{Illustration of topological (dis)similarities between Predicted-Aging, Predicted-NonAging, NotPredicted-Aging, and NotPredicted-NonAging genes in a given subnetwork by embedding their features into 2D space. This figure is for Berchtold-BioGRID; similar embedding trends hold for the other three subnetworks (Supplementary Figs. S8, S10, and S11). This is a zoomed-in visualization with outliers removed for simplicity (see Supplementary Fig. S9 for the full visualization). When mapping features into 2D space, we have tested $t$SNE and PCA and selected the visualization with the clearest pattern. This figure is generated using $t$SNE.}
  \label{fig:2d-berchtold-biogrid}
  \vspace{-0.1cm}
\end{figure}

Our predictive model(s) identify NotPredicted-Aging and NotPredicted-NonAging genes as \emph{not} being aging-related (Section \ref{sect:result-deep}). 
This implies that these genes' network features differ from network features of Predicted-Aging and Predicted-NonAging genes (which are predicted as aging-related by our network analysis). 
This is not surprising for NotPredicted-NonAging genes, as these are our negative control genes (Section \ref{sect:result-deep}). What is surprising is that NotPredicted-Aging genes, which are aging-related according to the ground truth data, seem to have network features that are more similar to those of NotPredicted-NonAging genes (because NotPredicted-NonAging genes are not predicted as aging-related) than those of Predicted-Aging and Predicted-NonAging (which are predicted as aging-related).  
Indeed, we confirm the network feature (dis)similarities between the different gene groups in the 2D visualization (Fig. \ref{fig:2d-berchtold-biogrid} and Supplementary Figs. S8-S11). 
This further supports postulation 3 that our inferred subnetworks or predictive models for their analyses might have failed to recognize all network ``signatures'' of GenAge aging-related genes, i.e., that they have failed to identify as aging-related the ``signatures'' of NotPredicted-Aging genes.

This result suggests a need for more advanced subnetwork inference or analysis methods. To infer our subnetworks, we relied on network propagation. A potential solution to improve subnetwork inference is to examine whether other types of integrative algorithms, such as kernel-, Bayesian-, or non-negative matrix factorization-based methods \cite{gligorijevic2015methods}, would result in subnetworks that yield higher prediction accuracy or uncover complementary aspects of aging-related knowledge. A potential solution to improve network analysis is to use an ensemble learning approach that may be able to integrate complementary aspects of different subnetworks \cite{sagi2018ensemble},  as suggested in Section \ref{sect:discuss-model}. Another potential solution is to use deep learning (e.g., graph convolutional networks) for dynamic network analysis \cite{pareja2020evolvegcn} to automatically learn and capture the differing types of network ``signatures'' of genes of interest. 

\section{Conclusions}
In this study, we have systematically evaluated whether using newer GE or PPIN data curated via more advanced biotechnologies to construct an aging-related subnetwork would improve the accuracy of predicting aging-related genes from the subnetworks. Unexpectedly, we have found that using newer GE or PPIN data does not help compared to using older data. In fact, the subnetwork inferred from both older Berchtold GE data and older HPRD PPIN data marginally outperforms the other subnetworks inferred from newer Berchtold GE data or newer HPRD PPIN data (or both). We have performed several analyses to try to explain this, without a clear answer. We have provided guidance on future directions towards the advancement of computational aging research, including a need for more complete aging-related ground truth data as well as more advanced subnetwork inference and analysis methods. 


\section{Methods} \label{sect:method}
\subsection{Data}  \label{sect:method-data}
\subsubsection{Considered aging-specific GE data} \label{sect:method-data-GE}
The two considered aging-specific GE data sets are from Berchtold \cite{berchtold2008gene} and GTEx \cite{gtex2015genotype}. Berchtold GE data encompasses 173 postmortem samples in brain from 55 individuals, which span 37 different ages from 20 to 99 years. GTEx GE data (version 8) encompasses 17,382 postmortem samples across 54 tissues from 175 individuals, which span six  age groups: 20-29, 30-39, 40-49, 50-59, 60-69, and 70-79. Of the 54 tissues from 31 organs, 13 tissues are from the brain. These are tissues that we focus on, for fair comparison to Berchtold data that is solely brain-related. The 13 tissues encompass 2,642 samples. 

\subsubsection{Considered entire context-unspecific PPIN data} \label{sect:method-data-PPIN}
The two considered entire context-unspecific PPINs are from HPRD \cite{berchtold2008gene} and BioGRID (version 4.4.197, the latest version at the time we started this study) \cite{oughtred2021biogrid}. We consider only the largest connected component of the two PPINs. The one from HPRD has 8,938 proteins and 35,900 interactions. The largest connected component of human physical interactions from BioGRID has 18,928 proteins and 484,146 interactions. 

\subsubsection{Inference of weighted dynamic aging-specific subnetworks} \label{sect:method-data-infer}

We infer multiple weighted dynamic aging-specific subnetworks, depending on which GE and PPIN data are used.  Berchtold-HPRD is the benchmark subnetwork from our previous study \cite{li2021improved}. Berchtold-HPRD was inferred using a method called NetWalk \cite{Komurov2010}, which was proved to be the best of two state-of-the-art network propagation algorithms for inferring aging-specific subnetworks in our previous study \cite{newaz2020improving}. For fairness, we use NetWalk to infer all other considered networks.

Namely, for an age (in case of Berchtold GE data) or age group (in case of GTEx GE data), NetWalk propagates gene expression levels onto the interactions in an entire context-unspecific PPIN via a biased random walk, and outputs a PPIN with two directed age-specific weights assigned to each interaction. That is, an interaction ($u, v$) has a weight from $u$ to $v$ and another weight from $v$ to $u$. The smaller of the two weights is selected for each interaction, yielding a weighted static age-specific subnetwork snapshot for the given age (group). This way, we create one static weighted subnetwork snapshot for each of the considered ages or age groups.

Given all $x$ weighted static age-specific subnetwork snapshots for the $x$ considered ages or age groups, we obtain the final dynamic weighted aging-specific subnetwork (for a given combination of GE and PPIN data) using an established two-step approach \cite{li2021improved}. 
First, to make the weights of the interactions comparable across snapshots, the weights are  normalized over all $x$ snapshots. Second, given the $x$  snapshots with normalized interaction weights, $x-1$ ``differential'' snapshots are created; a ``differential'' snapshot is created for each pair of consecutive normalized-weight  snapshots $i$ and $i+1$ ($i=1,2,3,...,x-1$), where for each interaction, its ``differential'' weight $w_{i,i+1}$ is defined as the relative weight change between the snapshots $i$ (i.e., $w_i$) and $i+1$ (i.e., $w_{i+1}$): $w_{i,i+1} = \frac{[w_{i+1} - W_i]\times 100}{[w_{i+1} + w_i]}$. The collection of the $x-1$ age-specific ``differential'' snapshots in the increasing order of ages or age groups forms the final weighted dynamic aging-specific subnetwork (for a given combination of GE and PPIN data).

Because we aim to evaluate whether subnetworks inferred using newer GE data or newer PPIN data (or both) outperform the benchmark subnetwork (Berchtold-HPRD) that is inferred using both older GE data and older PPIN data, we need to systematically test which  component (GE data or PPIN data) leads to potential performance improvement. To do so, in addition to Berchtold-HPRD, we infer three additional aging-specific subnetworks (also using NetWalk) by varying one data component at a time. That is, compared to Berchtold-HPRD, by changing the PPIN data component, we infer Berchtold-BioGRID; by changing the GE data component, we infer GTEx-HPRD; by changing both the GE and the PPIN data component, we infer GTEx-BioGRID. In addition, to examine whether grouping samples into individual (specific) ages vs. age groups yields performance difference  (Section \ref{sect:result-cv}), we group samples in Berchtold GE data into six age groups that match those in GTEx GE data (see above). Note that of the 173 samples in Berchtold GE data, we focus on all 119 samples that belong to one of the six age groups. Then, we combine this modified, age group-based Berchtold GE data with each of HPRD and BioGRID PPIN data using NetWalk to infer two additional subnetworks, referred to as Berchtold-HPRD-6 and Berchtold-BioGRID-6, respectively.  Table \ref{table:netstat} shows sizes of the six subnetworks.

\begin{table}[!htp]
    \centering
    \caption{The number of snapshots and size for each of the six considered weighted dynamic aging-specific subnetworks.}
    \vspace{-0.1cm}
    \label{table:netstat}
    \begin{tabular}{|l|c|c|c|}
        \hline
        Subnetwork          & \# of snapshots  & \# of nodes  & \# of edges   \\ \hline
        Berchtold-HPRD      &  37              & 8,938       &   35,900     \\ \hline
        Berchtold-BioGRID   &  37              & 18,928      &   484,146    \\ \hline
        GTEx-HPRD           &  6               & 8,938       &   35,900     \\ \hline
        GTEx-BioGRID        &  6               & 18,928      &   484,146    \\ \hline
        Berchtold-HPRD-6    &  6               & 8,938       &   35,900     \\ \hline
        Berchtold-BioGRID-6 &  6               & 18,928      &   484,146    \\ \hline
    \end{tabular}
\end{table}

Note that network propagation requires, for each age (group), expression levels for all genes in an entire context-unspecific PPIN. Also, because GE data encompasses multiple samples for each age (group), the expression level of a gene at a given age (group) needs to be somehow combined or normalized across these samples.  We achieve  these tasks as follows.

Berchtold GE data was curated using microarray technology (Affymetrix Hg-U133plus 2.0), containing expression information for 54,675 probes. Because nodes in our PPIN data have gene IDs, to integrate GE with PPIN data, we need to map probe IDs to gene IDs. To do this, we mimic our previous study \cite{newaz2020improving}. That is, by using the DAVID tool \cite{huang2009extracting}, we are able to map 48,724 of the 54,675 probes to 21,441 unique gene IDs. 
Next, while there are multiple samples (i.e., expression values) relevant to a gene at a given age, we need to obtain a single score value for each gene at each age. We do this by assigning to a given gene the average expression value of all of its  samples relevant to the given age. Moreover, we use the MAS 5.0 BioConductor package \cite{gentleman2004bioconductor} that calculates the $p$-value of whether a gene that is present in Berchtold GE data is statistically significantly expressed (i.e., active) at a given age. This enables us to assign expression values to those genes that are present in the considered PPIN but not in Berchtold GE data at a given age. We assume such genes to not be active at a given age. This is why we assign them the average expression value of all non-active  genes that are present in Berchtold GE data at the given age. For methodological details, see Supplementary Section S1.1.1 of \cite{faisal2014dynamic}.

GTEx GE data was curated using RNA-Seq techology. There are many methods proposed for normalizing RNA-Seq data. We use the Trimmed Mean of the M-values (TMM) \cite{robinson2010scaling} because \emph{(1)} TMM has been widely used for this purpose; and \emph{(2)} it was suggested by \cite{evans2018selecting, zhao2021tpm} that TMM typically performs better than  other existing normalization methods. In particular, TMM assumes that most of the genes are not deferentially expressed, and the batch effect across multiple tissues or genotypes can be removed. Briefly, TMM works as follows. First, the log2 fold change and absolute expression value of a sample against the reference sample are calculated. Then, those genes that have high fold changes or have large absolute expression values are trimmed, and the weighted average fold change per sample can be calculated. Finally, the read counts of genes are normalized by the weighted average fold change and the total number of samples in the GE data. For methodological details about TMM, see \cite{robinson2010scaling}. We use a BioConductor package called edgeR \cite{robinson2010edger} to obtain normalized GE data via TMM. Note that genes in GTEx GE data are represented as Ensembl IDs, and we convert these IDs to gene symbols using the DAVID tool. We remove from consideration a gene if its Ensembl ID shares more than one gene symbol. Unlike microarray-based GE data, we can not obtain $p$-value for whether a gene is statistically significantly expressed at a given age group due to the lack of reference data. So, for those genes that are present in the considered PPINs but not in GTEx GE data at a given age group, we assign these genes an expression value of zero.

\subsubsection{Aging- and non-aging-related gene labels} \label{sect:method-data-gt}
For fairness in comparing the considered subnetworks, we consider all 8,756 genes that are present in all of the subnetworks.  Supervised classification requires ground truth labels, i.e., knowledge of which of the considered genes are aging- vs.  non-aging-related. 

To define aging-related gene labels, as established \cite{li2021improved}, we rely on a confident ground truth data source, GenAge \cite{tacutu2017human}. Human genes included in GenAge are sequence orthologs of aging-related genes in model species. All aging-related genes in model species are experimentally validated. Their human homologs are included in GenAge if they are  supported by multiple pieces of literature and have aging-related phenotypical evidence in the human species. Of all 307 genes in GenAge, 277 of them are among the 8,756 considered genes. We denote these 277 genes as \underline{aging-related genes}. 

Next, we define non-aging-related genes from the $8,756 - 277 = 8,479$ genes that are not present in GenAge. To ensure that our non-aging-related labels are as confident as possible, we also rely on  five other aging-related ground truth data sets curated by \cite{jia2018analysis, lu2004gene, berchtold2008gene, simpson2011microarray}. That is, we remove all genes that are present in any of the five aging-related ground truth data sets from the 8,479 genes. This leaves 4,282 genes that we denote as \underline{non-aging-related genes}. We refer to the $277 + 4282 = 4559$ aging- and non-aging-related genes as \underline{ground truth labeled genes}.

\subsubsection{Aging-related pathways and GO terms} \label{sect:method-data-pathway}
We focus on five confident aging-related pathways, i.e., adenosine monophosphate-activated kinase (AMPK), Sirtuin 1 (SIRT1), mammalian target of rapamycin (mTOR),  phosphatidylinositol-3-kinase and protein kinase B (PI3K-AKT), and neurotrophin. Their relatedness to the aging process and lifespan was examined by numerous studies and was systematically reviewed by \cite{yu2021key}, as follows.

\begin{itemize}
    \item AMPK, SIRT1, and mTOR pathways are some of the key pathways that describe molecular mechanisms of the aging process. In particular, AMPK is considered as aging-related because it was found to control cell survival, growth, death, and autophagy, and it can regulate cellular homeostasis and resistance to stress \cite{morgunova2019age}. SIRT1 is considered as aging-related because it was found to play an important role in improving oxidative stress resistance of cells and inhibiting cell death \cite{brunet2004stress}, and hence is involved in many age-related diseases \cite{zhao2020sirtuins}. mTOR is considered aging-related because it functions as a sensor of intracellular energy and a central regulator of biological processes, including aging \cite{di2018intermittent, arriola2016rapamycin}.
    \item PI3K-AKT pathway is considered as aging-related because it was identified as the signaling pathway of several aging-related diseases: diabetic encephalopathy \cite{wang2018autophagy} and cancer \cite{porta2014targeting}.
    \item Neurotrophin pathway is considered as aging-related because it was identified as the key signaling pathway of aging-related Parkinson’s disease \cite{paudel2020emerging}.
\end{itemize}

We obtain the genes for these five pathways from the KEGG database \cite{kanehisa2021kegg}. We summarize the size of each pathway as in KEGG and as in our ground truth aging-related data in Table \ref{table:pathwaystat}. In addition, we consider aging-related GO terms (Supplementary Section S1.1).

\begin{table}[!ht]
    \centering
    \caption{The sizes of (numbers of genes in) the five aging-related pathways as available in KEGG and in our ground truth data. That is, the latter is the number of genes from a given KEGG pathway that are present in our 4,559 ground truth labeled genes.}
    \vspace{-0.1cm}
    \label{table:pathwaystat}
    \begin{tabular}{|l|c|c|}
        \hline
        Pathway       & Size in KEGG  & Size in our ground truth data  \\ \hline
        AMPK          & 113   & 38    \\ \hline
        SIRT1         & 89    & 36    \\ \hline
        mTOR          & 165   & 41    \\ \hline
        PI3K-AKT      & 354   & 179   \\ \hline
        Neurotrophin  & 119   & 57   \\ \hline
    \end{tabular}
\end{table}


\section*{Competing interests}
\vspace{-0.1cm}
There is NO competing interest.
\vspace{-0.2cm}
\section*{Author contributions statement}
\vspace{-0.1cm}
Q.L. and T.M. designed the study. Q.L. carried out all computational experiments. Q.L. and T.M. analyzed the results. K.N. helped provide guidance on how to process Berchtold and GTEx GE data and a part of the code needed to pre-process the data. Q.L. drafted the initial paper, and all authors contributed to the writing of the final version. All authors approved the manuscript. T.M. supervised all aspects of the study. 
\vspace{-0.2cm}
\section*{Acknowledgments}
\vspace{-0.1cm}
This work is supported by funds from the National Science Foundation (NSF CAREER CCF-1452795). 
\vspace{-0.2cm}

\bibliographystyle{unsrt}  
\bibliography{reference}

\begin{thebibliography}{10}

\bibitem{uyar2020single}
Bora Uyar et~al.
\newblock Single-cell analyses of aging, inflammation and senescence.
\newblock {\em Ageing Research Reviews}, 64:101156, 2020.

\bibitem{ferrucci2020measuring}
Luigi Ferrucci et~al.
\newblock Measuring biological aging in humans: A quest.
\newblock {\em Aging Cell}, 19(2):e13080, 2020.

\bibitem{rodriguez2011aging}
Sandra Rodr{\'\i}guez-Rodero et~al.
\newblock Aging genetics and aging.
\newblock {\em Aging and Disease}, 2(3):186, 2011.

\bibitem{liguori2018oxidative}
Ilaria Liguori et~al.
\newblock Oxidative stress, aging, and diseases.
\newblock {\em Clinical Interventions in Aging}, 13:757, 2018.

\bibitem{bolignano2014aging}
Davide Bolignano et~al.
\newblock The aging kidney revisited: a systematic review.
\newblock {\em Ageing Research Reviews}, 14:65--80, 2014.

\bibitem{paschos2012obesity}
Georgios~K Paschos et~al.
\newblock Obesity in mice with adipocyte-specific deletion of clock component
  arntl.
\newblock {\em Nature Medicine}, 18(12):1768--1777, 2012.

\bibitem{lu2004gene}
Tao Lu et~al.
\newblock Gene regulation and {DNA} damage in the ageing human brain.
\newblock {\em Nature}, 429(6994):883, 2004.

\bibitem{berchtold2008gene}
Nicole~C Berchtold et~al.
\newblock Gene expression changes in the course of normal brain aging are
  sexually dimorphic.
\newblock {\em Proceedings of the National Academy of Sciences},
  105(40):15605--15610, 2008.

\bibitem{simpson2011microarray}
Julie~E Simpson et~al.
\newblock Microarray analysis of the astrocyte transcriptome in the aging
  brain: relationship to {A}lzheimer's pathology and {APOE} genotype.
\newblock {\em Neurobiology of Aging}, 32(10):1795--1807, 2011.

\bibitem{tacutu2017human}
Robi Tacutu et~al.
\newblock Human {A}geing {G}enomic {R}esources: new and updated databases.
\newblock {\em Nucleic Acids Research}, 46(D1):D1083--D1090, 2017.

\bibitem{jia2018analysis}
Kaiwen Jia et~al.
\newblock An analysis of aging-related genes derived from the genotype-tissue
  expression project ({GTEx}).
\newblock {\em Cell Death Discovery}, 5(1):26, 2018.

\bibitem{emanuel2000makes}
Ezekiel~J Emanuel et~al.
\newblock What makes clinical research ethical?
\newblock {\em Jama}, 283(20):2701--2711, 2000.

\bibitem{fabris2017review}
Fabio Fabris et~al.
\newblock A review of supervised machine learning applied to ageing research.
\newblock {\em Biogerontology}, 18(2):171--188, 2017.

\bibitem{holtman2015induction}
Inge~R Holtman et~al.
\newblock Induction of a common microglia gene expression signature by aging
  and neurodegenerative conditions: a co-expression meta-analysis.
\newblock {\em Acta Neuropathologica Communications}, 3(1):31, 2015.

\bibitem{chen2014identifying}
Bolin Chen et~al.
\newblock Identifying protein complexes and functional modules — from static
  {PPI} networks to dynamic {PPI} networks.
\newblock {\em Briefings in Bioinformatics}, 15(2):177--194, 2014.

\bibitem{freitas2011data}
Alex~A Freitas et~al.
\newblock A data mining approach for classifying {DNA} repair genes into
  ageing-related or non-ageing-related.
\newblock {\em BMC Genomics}, 12(1):27, 2011.

\bibitem{fang2013classifying}
Yaping Fang et~al.
\newblock Classifying aging genes into {DNA} repair or non-{DNA} repair-related
  categories.
\newblock In {\em International Conference on Intelligent Computing}, pages
  20--29. Springer, 2013.

\bibitem{fabris2016extensive}
Fabio Fabris et~al.
\newblock An extensive empirical comparison of probabilistic hierarchical
  classifiers in datasets of ageing-related genes.
\newblock {\em IEEE/ACM Transactions on Computational Biology and
  Bioinformatics}, 13(6):1045--1058, 2016.

\bibitem{kerepesi2018prediction}
Csaba Kerepesi et~al.
\newblock Prediction and characterization of human ageing-related proteins by
  using machine learning.
\newblock {\em Scientific Reports}, 8(1):4094, 2018.

\bibitem{faisal2014dynamic}
Fazle~E Faisal and Tijana Milenkovi{\'c}.
\newblock Dynamic networks reveal key players in aging.
\newblock {\em Bioinformatics}, 30(12):1721--1729, 2014.

\bibitem{Elhesha2019}
Rasha Elhesha et~al.
\newblock Identification of co-evolving temporal networks.
\newblock {\em BMC Genomics}, 20(434), 2019.

\bibitem{li2020supervisedTCBB}
Qi~Li and Tijana Milenkovi{\'c}.
\newblock Supervised prediction of aging-related genes from a context-specific
  protein interaction subnetwork$\dagger$.
\newblock {\em IEEE/ACM Transactions on Computational Biology and
  Bioinformatics}, 2021.

\bibitem{li2021improved}
Qi~Li et~al.
\newblock Improved supervised prediction of aging-related genes via weighted
  dynamic network analysis.
\newblock {\em BMC Bioinformatics}, 22(1):1--26, 2021.

\bibitem{li2019supervisedBIBM}
Qi~Li and Tijana Milenkovi{\'c}.
\newblock Supervised prediction of aging-related genes from a context-specific
  protein interaction subnetwork.
\newblock {\em IEEE International Conference on Bioinformatics and Biomedicine
  (BIBM)}, pp:130--137, 2019.

\bibitem{newaz2020improving}
Khalique Newaz and Tijana Milenkovi\'{c}.
\newblock Inference of a dynamic aging-related biological subnetwork via
  network propagation.
\newblock {\em IEEE/ACM Transactions on Computational Biology and
  Bioinformatics}, 2020.

\bibitem{Komurov2010}
Kakajan Komurov et~al.
\newblock Use of data-biased random walks on graphs for the retrieval of
  context-specific networks from genomic data.
\newblock {\em {PLOS} Computational Biology}, 6(8):1--10, 2010.

\bibitem{leiserson2015pan}
Mark~DM Leiserson et~al.
\newblock Pan-cancer network analysis identifies combinations of rare somatic
  mutations across pathways and protein complexes.
\newblock {\em Nature Genetics}, 47(2):106--114, 2015.

\bibitem{keshava2008human}
TS~Prasad et~al.
\newblock Human {P}rotein {R}eference {D}atabase—2009 update.
\newblock {\em Nucleic Acids Research}, 37(suppl\_1):D767--D772, 2009.

\bibitem{gtex2015genotype}
{GTEx Consortium}.
\newblock The {Genotype-Tissue Expression} ({GTEx}) pilot analysis:
  {Multitissue} gene regulation in humans.
\newblock {\em Science}, 348(6235):648--660, 2015.

\bibitem{oughtred2021biogrid}
Rose Oughtred et~al.
\newblock The {BioGRID} database: A comprehensive biomedical resource of
  curated protein, genetic, and chemical interactions.
\newblock {\em Protein Science}, 30(1):187--200, 2021.

\bibitem{kanehisa2021kegg}
Minoru Kanehisa et~al.
\newblock {KEGG}: integrating viruses and cellular organisms.
\newblock {\em Nucleic Acids Research}, 49(D1):D545--D551, 2021.

\bibitem{gene2021gene}
{Gene Ontology Consortium}.
\newblock {The Gene Ontology resource: enriching a GOld mine}.
\newblock {\em Nucleic Acids Research}, 49(D1):D325--D334, 2021.

\bibitem{yu2021key}
Mengdi Yu et~al.
\newblock Key signaling pathways in aging and potential interventions for
  healthy aging.
\newblock {\em Cells}, 10(3):660, 2021.

\bibitem{lazareva2021limits}
Olga Lazareva et~al.
\newblock On the limits of active module identification.
\newblock {\em Briefings in Bioinformatics}, 2021.

\bibitem{newaz2020network}
Khalique Newaz et~al.
\newblock Network analysis of synonymous codon usage.
\newblock {\em Bioinformatics}, 36(19):4876--4884, 2020.

\bibitem{hulovatyy2016scout}
Yuriy Hulovatyy and Tijana Milenkovi{\'c}.
\newblock {SCOUT}: simultaneous time segmentation and community detection in
  dynamic networks.
\newblock {\em Scientific Reports}, 6(1):1--11, 2016.

\bibitem{bronikowski2011aging}
Anne~M Bronikowski et~al.
\newblock Aging in the natural world: comparative data reveal similar mortality
  patterns across primates.
\newblock {\em Science}, 331(6022):1325--1328, 2011.

\bibitem{faisal2014global}
Fazle~Elahi Faisal et~al.
\newblock Global network alignment in the context of aging.
\newblock {\em IEEE/ACM Transactions on Computational Biology and
  Bioinformatics}, 12(1):40--52, 2014.

\bibitem{gligorijevic2015methods}
Vladimir Gligorijevi{\'c} and Nata{\v{s}}a Pr{\v{z}}ulj.
\newblock Methods for biological data integration: perspectives and challenges.
\newblock {\em Journal of the Royal Society Interface}, 12(112):20150571, 2015.

\bibitem{sagi2018ensemble}
Omer Sagi and Lior Rokach.
\newblock Ensemble learning: {A} survey.
\newblock {\em Wiley Interdisciplinary Reviews: Data Mining and Knowledge
  Discovery}, 8(4):e1249, 2018.

\bibitem{pareja2020evolvegcn}
Aldo Pareja et~al.
\newblock Evolvegcn: {E}volving graph convolutional networks for dynamic
  graphs.
\newblock In {\em Proceedings of the AAAI Conference on Artificial
  Intelligence}, volume~34, pages 5363--5370, 2020.

\bibitem{huang2009extracting}
Da~Wei Huang et~al.
\newblock Extracting biological meaning from large gene lists with {DAVID}.
\newblock {\em Current Protocols in Bioinformatics}, 27(1):13--11, 2009.

\bibitem{gentleman2004bioconductor}
Robert~C Gentleman et~al.
\newblock {B}ioconductor: open software development for computational biology
  and bioinformatics.
\newblock {\em Genome Biology}, 5(10):1--16, 2004.

\bibitem{robinson2010scaling}
Mark~D Robinson and Alicia Oshlack.
\newblock A scaling normalization method for differential expression analysis
  of {RNA-seq} data.
\newblock {\em Genome Biology}, 11(3):1--9, 2010.

\bibitem{evans2018selecting}
Ciaran Evans et~al.
\newblock Selecting between-sample {RNA-Seq} normalization methods from the
  perspective of their assumptions.
\newblock {\em Briefings in Bioinformatics}, 19(5):776--792, 2018.

\bibitem{zhao2021tpm}
Yingdong Zhao et~al.
\newblock {TPM, FPKM}, or normalized counts? a comparative study of
  quantification measures for the analysis of {RNA-seq} data from the {NCI}
  patient-derived models repository.
\newblock {\em Journal of Translational Medicine}, 19(1):1--15, 2021.

\bibitem{robinson2010edger}
Mark~D Robinson et~al.
\newblock {edgeR}: a {Bioconductor} package for differential expression
  analysis of digital gene expression data.
\newblock {\em Bioinformatics}, 26(1):139--140, 2010.

\bibitem{morgunova2019age}
Galina~V Morgunova and Alexander~A Klebanov.
\newblock Age-related {AMP}-activated protein kinase alterations: From cellular
  energetics to longevity.
\newblock {\em Cell Biochemistry and Function}, 37(3):169--176, 2019.

\bibitem{brunet2004stress}
Anne Brunet et~al.
\newblock Stress-dependent regulation of {FOXO} transcription factors by the
  {SIRT1} deacetylase.
\newblock {\em Science}, 303(5666):2011--2015, 2004.

\bibitem{zhao2020sirtuins}
Lijun Zhao et~al.
\newblock Sirtuins and their biological relevance in aging and age-related
  diseases.
\newblock {\em Aging and Disease}, 11(4):927, 2020.

\bibitem{di2018intermittent}
Andrea Di~Francesco et~al.
\newblock Intermittent {mTOR} inhibition reverses kidney aging in old rats.
\newblock {\em The Journals of Gerontology Series A: Biological Sciences and
  Medical Sciences}, 73(7):843, 2018.

\bibitem{arriola2016rapamycin}
Sebastian~I Arriola~Apelo and Dudley~W Lamming.
\newblock Rapamycin: an {InhibiTOR} of aging emerges from the soil of easter
  island.
\newblock {\em Journals of Gerontology Series A: Biomedical Sciences and
  Medical Sciences}, 71(7):841--849, 2016.

\bibitem{wang2018autophagy}
Beiyun Wang et~al.
\newblock Autophagy of macrophages is regulated by {PI3k/Akt/mTOR} signalling
  in the development of diabetic encephalopathy.
\newblock {\em Aging (Albany NY)}, 10(10):2772, 2018.

\bibitem{porta2014targeting}
Camillo Porta et~al.
\newblock Targeting {PI3K/Akt/mTOR} signaling in cancer.
\newblock {\em Frontiers in Oncology}, 4:64, 2014.

\bibitem{paudel2020emerging}
Yam~Nath Paudel et~al.
\newblock Emerging neuroprotective effect of metformin in parkinson’s
  disease: A molecular crosstalk.
\newblock {\em Pharmacological Research}, 152:104593, 2020.

\end{thebibliography}

\end{document}